# Janus monolayer TaNF: a new ferrovalley material with large valley splitting and tunable magnetic properties

Guibo Zheng (郑贵博)[1], Shuixian Qu (蘧水仙)[1], Wenzhe Zhou (周文哲)[1,*] and Fangping Ouyang (欧阳方平) [1,2,3,*]

[1]*School of Physics and Electronics, and Hunan Key Laboratory for Super-Microstructure and Ultrafast Process, and Hunan Key Laboratory of Nanophotonics and Devices, Central South University, Changsha 410083, People's Republic of China*

[2]*School of Physics and Technology, Xinjiang University, Urumqi 830046, People's Republic of China*

[3]*State Key Laboratory of Powder Metallurgy, and Powder Metallurgy Research Institute, Central South University, Changsha 410083, People's Republic of China*

## Abstract

Materials with large intrinsic valley splitting and high Curie temperature are a huge advantage for studying valleytronics and practical applications. In this work, using first-principles calculations, a new Janus TaNF monolayer is predicted to exhibit excellent piezoelectric properties and intrinsic valley splitting, resulting from the spontaneous spin polarization, the spatial inversion symmetry breaking and strong spin-orbit coupling (SOC). TaNF is also a potential two-dimensional (2D) magnetic material due to its high Curie temperature and huge magnetic anisotropy energy. The effective control of the band gap of TaNF can be achieved by biaxial strain, which can transform TaNF monolayer from semiconductor to semi-metal. The magnitude of valley splitting at the CBM can be effectively tuned by biaxial strain due to the changes of orbital composition at the valleys. The magnetic anisotropy energy (MAE) can be manipulated by changing the energy and occupation (unoccupation) states of *d* orbital compositions through biaxial strain. In addition, Curie temperature reaches 373 K under only -3% biaxial strain, indicating that Janus TaNF monolayer can be used at high temperatures for spintronic and valleytronic devices.

Keywords: Janus, Valley splitting, Curie temperature, Magnetic anisotropy energy, First-principles calculations

*Corresponding author. E-mail address: csuzwz22@csu.edu.cn (W. Zhou) and ouyangfp06@tsinghua.org.cn (F. Ouyang)

Table of Contents

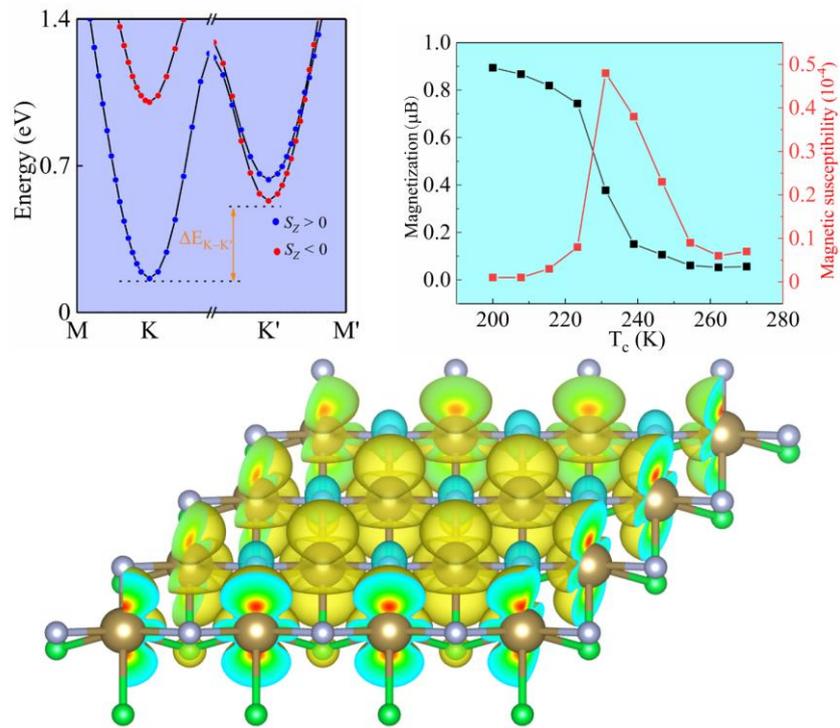

A huge valley splitting and high Curie temperature are found in a new ferrovalley semiconductor of Janus TaNF monolayer.

# 1. Introduction

The research of two-dimensional materials has promoted the rapid development of valleytronics[1, 2]. Due to the broken inversion symmetry and the strong spin-orbit coupling of transition metals, monolayer transition metal dichalcogenides (TMDs) have degenerate but not equivalent valley at K point and K' point in reciprocal space, which is applicable to future information storage and logic operation[3-5]. However, the most of valley materials are unusual of spontaneous valley splitting because they possess time reversal symmetry, which hinders their potential applications in valleytronics. Previous research suggests that lifting valley degeneracy can be achieved by various external engineering methods, including Optical Stark effect using ultrafast laser pumping[6, 7], magnetic atomic doping[8, 9] and the magnetic proximity effect[10, 11]. Although these methods can lift valley degeneracy and realize valley splitting, it is difficult for their practical applications. Therefore, searching for 2D materials with intrinsic valley splitting will not only be of great significance in exploring valley physics, but also beneficial to the practical application of valleytronic devices.

The ferrovalley material[12] with spontaneous valley splitting and intrinsic ferromagnetism are considered to have great potential for developing efficient spintronic nanodevices. The ferrovalley material have been found, such as GeSe[13] VSe$_2$[14-16], MnPS3[17] CuMP$_2$X$_6$(M=Cr, V X=S, Se)[18], GdX$_2$ (X=Br, Cl)[19] , YX$_2$(X=I, Br, and Cl)[20], FeCl$_2$[21-22]and Janus-VClBr[23]. Large perpendicular magnetic anisotropy can stabilize the orientation of the magnetic moment and form a long-range magnetic order. Therefore, it is vital to seek the 2D ferrovalley materials with considerable valley splitting, large magnetic anisotropy energy (MAE) and high Curie temperature (Tc) simultaneously for the development of valleytronic and applications for the integration of various electronic functions[24].

In this paper, by using first-principles calculations and Monte Carlo (MC) simulations[25], Janus TaNF monolayer are identified as excellent piezoelectric properties owing to big piezoelectric constant ($d_{31}$ = 0.33 pm/V), the promising 2D ferrovalley due to considerable valley splitting (370 meV) and 2D ferromagnetic (FM) semiconductor because of the large magnetic anisotropy energy up to 4.8 meV and high Tc beyond 220 K. In addition, the valley splitting, MAE and Tc can be effectively manipulated by biaxial strain, indicating that Janus TaNF monolayer is a very promising 2D ferrovalley and FM material for future integrated spintronic and valleytronic nanodevices.

## 2. Computational details

All structural optimization and electronic structure calculations were performed using the projector augmented wave (PAW)[26] method through the Vienna ab initio simulation package (VASP)[27]. For exchange and correlation interactions, the generalized gradient approximation (GGA) of Perdew-Burke-Ernzerhof functional (PBE)[28, 29] was treated. A vacuum region along the z direction was set to 20 Å so that the interaction between repeated slabs can be ignored. The energy cutoff of 550 eV and a 15 × 15 × 1 Γ-point centered grid were adopted. The energy difference between the two adjacent steps was less than $10^{-6}$ eV and atomic positions were fully relaxed until the maximum force on each atom was less than 0.01 eV/Å. SOC is considered in band structure calculations. we applied the PBE + U method [30] with $U_{eff} = (U - J) =$ 3.0 eV for $d$ orbital of Ta atoms to deal with strong correlation effects into account, where the $U_{eff}$ was determined using the linear response method[31-33], which have been widely adopted in previous research[34]. Phonon dispersion calculation was based on a 6 × 6 × 1 supercell by using the PHONOPY code [35] interfaced with the density-functional perturbation theory. A 5 × 5 × 1 supercell was adopted in ab initio molecular dynamic (AIMD) simulations[36] with a canonical ensemble and a Nosé thermostat. Piezoelectric coefficient and elastic constants are calculated by density functional perturbation theory (DFPT) methods, which have been widely reported in previous research[37, 38]. The Curie temperature (Tc) is simulated by using MC simulations based on the Heisenberg model.

## 3. Results and discussion

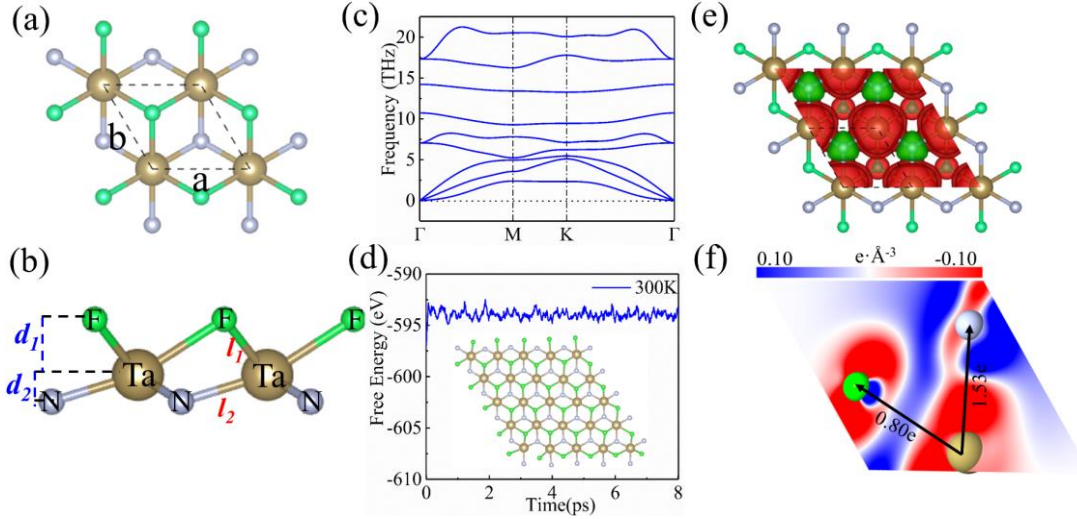

Fig. 1. (a) Top and (b) side views of the crystal structure of the Janus TaNF monolayer. The bond length of Ta-F is longer than that of Ta-N ($l_1 > l_2$), thus the distance between different sublayers $d_1$ is longer than $d_2$. (c) Phonon dispersion and (d) variation of total free energy with time during the ab initio molecular dynamics simulation (AIMD) for Janus TaNF monolayer at 300 K. The inset in (d) corresponds the snapshot taken from the end of the simulation. (e) Top views of the spin-polarized charge density of Janus TaNF monolayer. The spin-up and spin-down charge are indicated by red and green isosurfaces. (f) The differential charge density of F-Ta-N cross section, where the red and blue represent depletion and accumulation of electrons respectively.

The atomic structure of the Janus TaNF monolayer is shown in Fig. 1(a)-(b), which displays a hexagonal lattice. The bond lengths of Ta-F and Ta-N are different, which means the breaking of mirror symmetry. The Ta atoms are bonded with three N and three F atoms on two sides. The point group is $C_{3v}$ for Janus TaNF monolayer, and the structural parameters of Janus TaNF monolayer is listed in Table I.

The phonon dispersion along the high symmetry points in the Brillouin zone are calculated. No imaginary frequency appears in the whole Brillouin zone for Janus TaNF monolayer [Fig. 1(c)], which confirms its dynamical stability. The thermal stability of Janus monolayer is verified by performing ab initio molecular dynamics simulation (AIMD) simulations at 300 K. As shown in Fig. 1(d), the free energy of the 5 × 5× 1 supercell fluctuates within a certain range, and final structure after 8 ps shows no obvious distortion during the simulation process, indicating that Janus TaNF monolayer are thermally stable at room temperature. In order to certificate the mechanical stability of Janus TaNF monolayer, two independent elastic constants $C_{11}$ and $C_{12}$ of the hexagonal crystal systems based on density functional perturbation theory (DFPT) methods are calculated. As shown in Table I, $C_{11} = C_{22} = 102.9$ N/m and $C_{12} = 46.6$ N/m for Janus TaNF monolayer. Obviously, the elastic constants obey the Born Huang criteria ($C_{11}C_{12} - C_{12}^2 > 0$ and $C_{11} > 0$).

Table II represents the calculation results of $e_{11}$, $e_{31}$, $d_{11}$, and $d_{31}$ piezoelectric coefficients through DFPT methods for Janus TaNF monolayer. Janus TaNF monolayer attains the $e_{11}$ of $3.63 \times 10^{-10}$ C/m, and $d_{11}$ of 6.27 pm/V, which is higher than h-BN ($d_{11}$ = 0.60 pm/V), MoS$_2$ ($d_{11}$ = 3.73 pm/V) and MoSe$_2$ ($d_{11}$ = 4.72 pm/V) [39]. Compared with monolayer TMDs with only in-plane piezoelectricity, the broken inversion symmetry in Janus TaNF monolayer induce an out-of-plane dipole moment that cause out-of-plane piezoelectric properties. Janus TaNF monolayer possesses $e_{31}$ of $0.64 \times 10^{-10}$ C/m, and $d_{31}$ of 0.33 pm/V reveals the excellent out-of-plane piezoelectric properties, comparing Janus MoSSe ($d_{31}$ = 0.02 pm/V), MoSTe ($d_{31}$ = 0.028 pm/V)[40], and In$_2$SeTe ($d_{31}$ = 0.15 pm/V)[37].

Table I. Structural parameters and band gap of Janus TaNF monolayer. The lattice constant ($a = b$), the bond length of Ta-F($l_1$) and Ta-N ($l_2$), the distance between the sublayers of Ta, F($d_1$) and Ta, N ($d_2$) and elastic constants ($C_{11}$ and $C_{12}$) are shown. The band gap of Janus TaNF monolayer within PBE + SOC are also shown.

|      | $a, b$ (Å) | $l_1$ (Å) | $l_2$ (Å) | $d_1$ (Å) | $d_2$ (Å) | $C_{11}$(N/m) | $C_{12}$(N/m) | $E_g$(eV) |
|------|-----------|-----------|-----------|-----------|-----------|---------------|---------------|-----------|
| TaNF | 3.355     | 2.307     | 2.028     | 1.248     | 0.600     | 102.9         | 46.6          | 0.246     |

Table II. Piezoelectric coefficients ($e_{11}$, $e_{31}$, $d_{11}$, and $d_{31}$) of Janus TaNF monolayer, along with the ones of some typical TMDs 2D materials.

|             | $e_{11}$ (10$^{-10}$ C/m) | $e_{31}$ (10$^{-10}$ C/m) | $d_{11}$ (pm/V) | $d_{31}$ (pm/V) |
|-------------|---------------------------|---------------------------|-----------------|-----------------|
| TaNF        | 3.63                      | 0.64                      | 6.27            | 0.33            |
| MoS$_2$[39] | 3.64                      |                           | 3.73            |                 |
| MoSe$_2$[39]| 3.92                      |                           | 4.72            |                 |
| MoSSe[40]   | 3.74                      | 0.032                     | 3.76            | 0.02            |
| MoSTe[40]   | 4.53                      | 0.038                     | 5.04            | 0.028           |

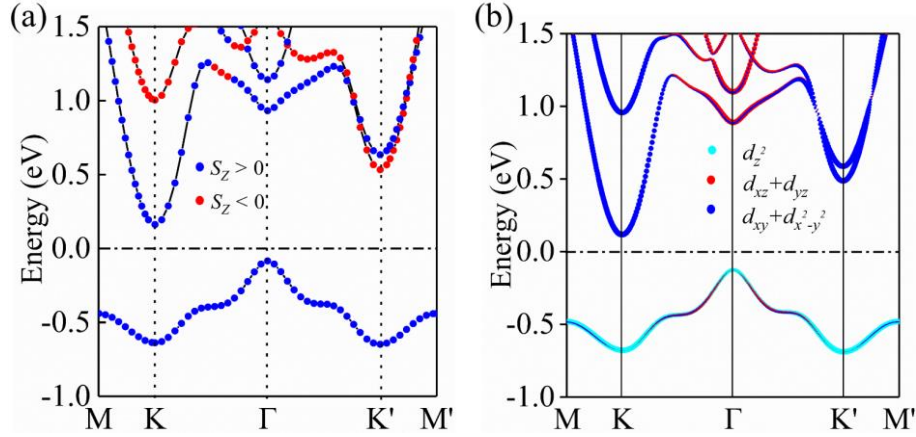

Fig. 2. (a) The spin projected band with SOC and (b) orbital projected band structure of Janus TaNF monolayer. The blue and red solid dots correspond to spin-up and spin-down of *z* direction.

The outer shell electron configuration of the Ta atom is $5d^36s^2$ and half-filled $5d$ orbital electrons result in a magnetic moment of 1 $\mu_B$/unitcell in Janus TaNF monolayer. In Fig. 1(e), the spin-up electrons mainly are distributed near the Ta atoms and spin-down electrons are induced around N atoms. The differential charge density of Janus TaNF monolayer is shown in Fig. 1(f), Ta atom transfer 1.53 and 0.80 electrons to N and F atoms, respectively. As shown in Fig. 2(a), Janus TaNF monolayer is semiconductors with indirect band gap of 0.246 eV considering spin-orbit coupling, the valence band maximum (VBM) and the conduction band minimum (CBM) are composed of spin-up electrons. Since the broken inversion symmetry and the exchange interaction of electrons, the energies of the K and K' valley electrons with inverse momentum are split under the SOC effect. The valley splitting at the bottom conduction band of Janus TaNF monolayer reaches 370 meV, which is much larger than that of $CrS_2$ (68 meV)[41], $VSe_2$ (78 meV)[42], CrOBr (112 meV)[43] and $VTe_2$ (157 meV)[44] in the previous reports. Fig. 2(b) shows the projected band structure of Janus TaNF monolayer. It is found that the K (K') valley of conduction band is mainly composed of the in-plane $d_{x^2-y^2}$ and $d_{xy}$ orbitals, while the K (K') point of valence band is majority constituted of the out-of-plane $d_{z^2}$ orbital, which explains why valley splitting occurs in the conduction band.

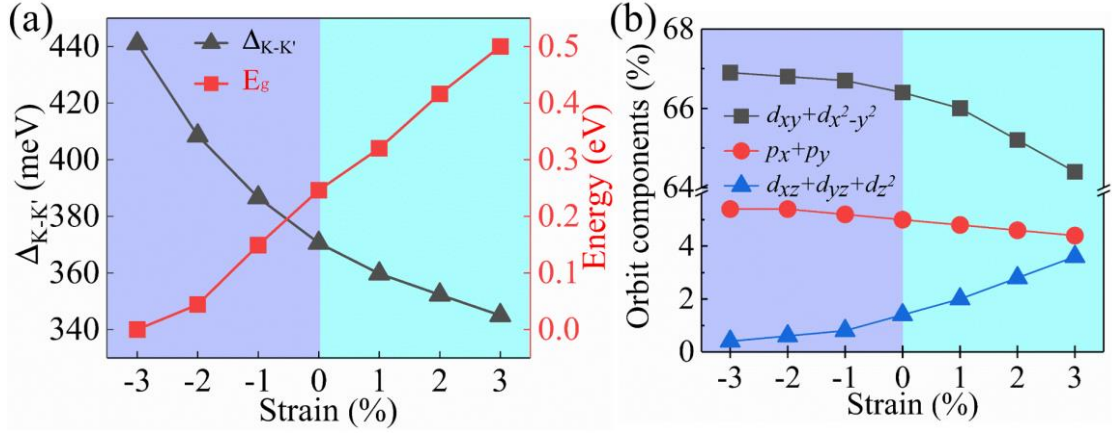

Fig. 3. (a) Dependence of valley splitting ($\Delta_{K-K'}$) and band gap ($E_g$) as functions of biaxial strain in the Janus TaNF monolayer. (b) The orbital components of K valley as functions of biaxial strain.

Fig. 3(a) shows the quantitative variation of valley splitting ($\Delta_{K-K'}$) and band gap ($E_g$) with biaxial strain ($\Delta a/a_0$) from -3% to 3%. With only -3% biaxial strain applied, the valley splitting will increase from 370 meV to 441 meV and band gap will decrease from 0.42 eV to 0 eV. In addition, Janus TaNF monolayers changes from semiconductor to semi-metal when the biaxial strain exceeds 3%. The valley splitting will decrease to 352 meV and the band gap will decrease to 0.50 eV under -3% biaxial strain. According to our previous research work[10], It can be seen from the formula that valley splitting is positively correlated with orbital angular momentum. We calculated the orbital components of CBM of K valley of Janus TaNF monolayers in Fig. 3(b). As shown in Fig. 3(b), with the increase of the compressive biaxial strain, the composition of $d_{xy}$ and $d_{x^2-y^2}$ orbitals increase, which contributes to the splitting of in-plane spin, and the composition of $d_{xz}$, $d_{yz}$ and $d_{z^2}$ orbitals decrease, which contributes to the splitting out-of-plane spin.

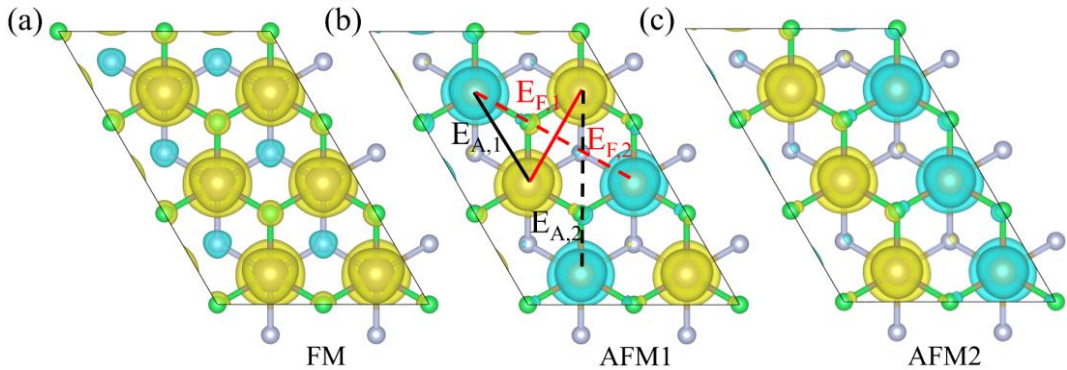

Fig. 4. Atomic arrangement of magnetic moments for (a) ferromagnetic and (b, c) antiferromagnetic

orders of Janus TaNF monolayer. (b) The schematic plot of magnetic bond energies $E_{F,1}$ and $E_{A,1}$ between the nearest Ta-Ta moments, and $E_{F,2}$ and $E_{A,2}$ between the next-nearest Ta-Ta moments. The spin-up and spin-down charge are indicated by yellow and green isosurfaces.

In order to determine the magnetic ground states of Janus TaNF monolayer, three possible magnetic configurations within the 3 × 2 × 1 supercell including the ferromagnetic (FM), antiferromagnetic (AFM1), and collinear antiferromagnetic (AFM2) spin arrangements are considered as depicted in the Fig. 3. The corresponding total energy of the three magnetic configurations offer the estimation of the exchange interaction parameters between the nearest-neighbor couplings $J_1$ and the next-nearest couplings $J_2$. The magnetic exchange coupling for the three magnetic configurations can be evaluated by the spin Heisenberg model.

$$H = E_0 + J_1 \sum_{<ij>} S_i \cdot S_j + J_2 \sum_{\ll ij \gg} S_i \cdot S_j \qquad (1)$$

where $J_1$ and $J_2$ represent the nearest-neighbor and the next-neighbor exchange coupling parameters, respectively, $S_i$ ($S_j$) is the unit vector of direction of the local magnetic moment at site $i$ ($j$). The constant $E_0$ includes all spin-independent interactions. To obtain the values of $J_1$ ($J_2$), one needs to evaluate the energy difference between a pair of nearest (next- nearest) Ta-Ta moments in parallel $E_{F,1}$ ($E_{F,2}$) and antiparallel $E_{A,1}$ ($E_{A,2}$) alignments,

$$2S^2 J_1 = E_{F,1} - E_{A,1} \qquad (2)$$

$$2S^2 J_2 = E_{F,2} - E_{A,2} \qquad (3)$$

The total energy for the Janus TaNF monolayer with FM, AFM1 and AFM2 ordering can be expressed by the following equations:

$$E_{FM} - E_{NM} = 3E_{F,1} + 3E_{F,2} \qquad (4)$$

$$E_{AFM1} - E_{NM} = E_{F,1} + 2E_{A,1} + 2E_{F,2} + E_{A,2} \qquad (5)$$

$$E_{AFM2} - E_{NM} = E_{F,1} + 2E_{A,1} + E_{F,2} + 2E_{A,2} \qquad (6)$$

we obtain the exchange interaction parameters $J_1$ and $J_2$ for Janus TaNF monolayer by solving the above equations with calculated total energy of the related spin states. The calculated exchange coupling parameters $J_1$ (-10.740 meV) and $J_2$ (-0.203 meV) are negative, which mean the ferromagnetic coupling between two Ta atoms in the nearest and the next-nearest shells.

MAE is defined by the energy difference between in-plane and out-of-plane ferromagnetic states and can be expressed as MAE = $E_x - E_z$, where $E_x$ and $E_z$ indicate the energy per unit cell with in-plane and out-of-plane ferromagnetic direction, respectively. The MAE (4.857 meV) is positive, which suggest that Janus TaNF monolayer possess a ferromagnetic ground state with out-of-plane magnetization.

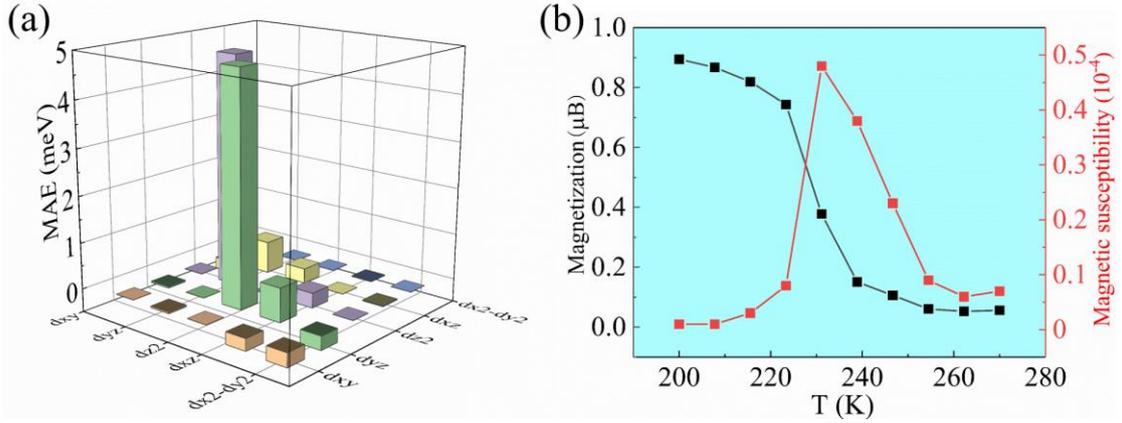

Fig. 5. (a) The *d*-orbital-projected-MAE of Ta atom for Janus TaNF monolayer. (b) Temperature variation of the magnetic moment and magnetic susceptibility for the Janus TaNF monolayer.

It is well known that a large uniaxial magnetic anisotropy can stabilize the orientation of the magnetic moment and form a long-range magnetic order under a finite temperature. So, large MAE is of great significance for the application of 2D magnetic materials. The MAE from SOC can be evaluated by the second-order perturbation. The MAE can be evaluated as MAE = $E_{up\text{-}up} + E_{up\text{-}down}$ under considering the interactions of spin polarizated states, which are written as,

$$E_{up\text{-}up} = \xi^2 \sum_{o^+ u^+} \frac{\left|\left\langle o^+ \left| L_z \right| u^+ \right\rangle\right|^2 - \left|\left\langle o^+ \left| L_x \right| u^+ \right\rangle\right|^2}{\varepsilon_u - \varepsilon_o} \qquad (7)$$

$$E_{up\text{-}down} = \xi^2 \sum_{o^+ u^-} \frac{\left|\left\langle o^+ \left| L_z \right| u^- \right\rangle\right|^2 - \left|\left\langle o^+ \left| L_x \right| u^- \right\rangle\right|^2}{\varepsilon_u - \varepsilon_o}, \qquad (8)$$

where o (u) denotes the occupied (unoccupied) states, and $L_z$, $L_x$ are the angular momentum operators. The SOC constant is represented by $\xi$. The energy $\varepsilon_u$ and $\varepsilon_o$ stands for the energy of unoccupied and occupied states, respectively. For the angular momentum matrix elements contributed by *d*-orbitals of Ta atom, there are five nonzero elements $\langle d_{xy}|L_x|d_{xz}\rangle$, $\langle d_{xy}|L_z|d_{x^2-y^2}\rangle$, $\langle d_{yz}|L_x|d_{z^2}\rangle$, $\langle d_{yz}|L_z|d_{xz}\rangle$, $\langle d_{xy}|L_x|d_{xz}\rangle$ and $\langle d_{yz}|L_x|d_{x^2-y^2}\rangle$. The contributions from different *d*-orbitals of Ta atom to MAE are studied in Janus TaNF monolayer. As shown in Fig. 6(d), the states near Fermi level are mainly from the $d_z^2$ and $d_{yz}$ orbitals, Therefore, the MAE mainly comes from $\langle d_{yz}|L_x|d_{z^2}\rangle$ since the absolute value of MAE is inversely proportional to the energy difference ($\varepsilon_u$ - $\varepsilon_o$), as described in Equation (7) and (8). Both occupied and unoccupied $d_{yz}$ and $d_z^2$ orbitals possess spin-up channels and cause to positive MAE.

To accurately estimate the Curie temperature, the Monte Carlo simulations for the magnetizations as functions of the temperature was adopted. The curve of magnetic moment and magnetic susceptibility versus temperature for Janus TaNF monolayer is shown in Fig. 5(b). It is found that the magnetic moment and magnetic susceptibility of the system decreases at 231 K. Therefore, the Curie temperature of Janus TaNF monolayer exceed those of many previously studied materials, such, $FeCl_2$ [45], $CrI_3$ [46] and $Cr_2Se_3$ [47].

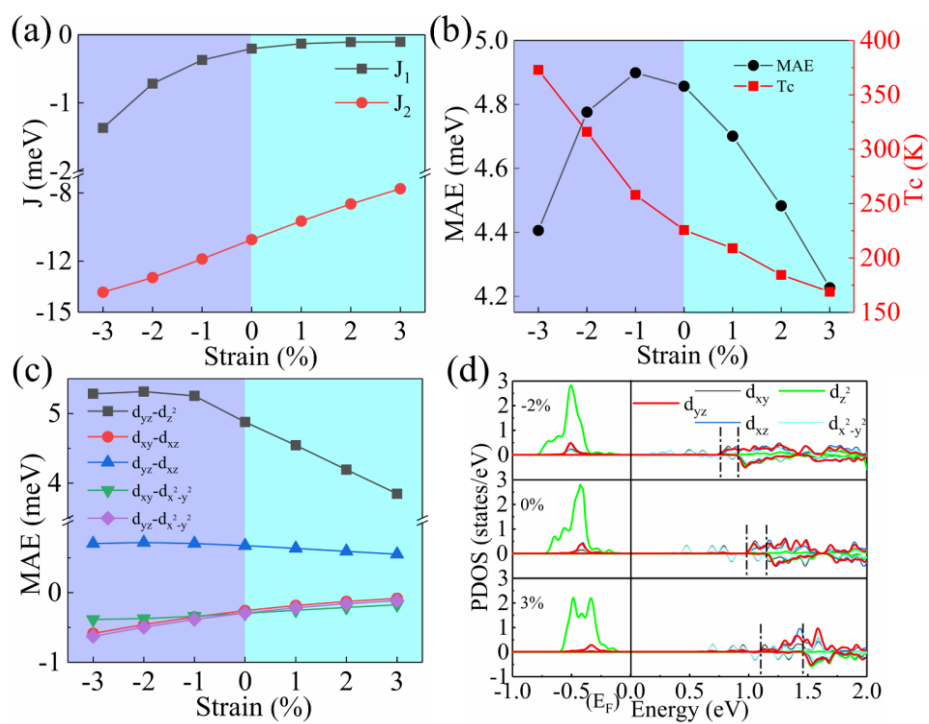

Fig. 6. The variations of exchange interaction parameters $J_1$ and $J_2$ (a), MAE and Curie temperature Tc (b) and projected orbital coupling matrix elements of Ta atoms (c) of the Janus TaNF monolayer as a function of the biaxial strain. (d) The d orbital PDOS near the Fermi level of Ta atom of Janus TaNF monolayer under a −2%, 0% and 3% biaxial strain. The black solid line represents Fermi level, and the two vertical black dashed lines show the bottom of spin-up and spin-down unoccupied states.

The change curves of exchange constants $J_1$ and $J_2$ with biaxial strain are shown in Fig. 6(a). Biaxial strain can regulate Curie temperature by tuning $J_1$ and $J_2$. It is indicated that $J_1$ and $J_2$ decrease with increasing biaxial compressive strain. So, $T_C$ monotonously increases with increasing strain from 3% to -3% and researches a maximum value of 373 K under -3% biaxial strain, indicating that Janus TaNF monolayer is a promising 2D magnetic material.

The change of MAE and projected orbital coupling matrix elements of Janus TaNF monolayer as a function of biaxial strain are shown Fig. 6(b) and Fig. 6(c), respectively. As the tensile strain increases, the decrease of the MAE of the Janus TaNF monolayer is mainly due to the decrease of $\langle d_{yz}|L_x|d_{z^2}\rangle$. With the increase of compressive strain, the MAE first increases and then decreases (-1% strain reaches the maximum) because $\langle d_{yz}|L_x|d_{z^2}\rangle$ first increases rapidly and then increases slowly while $\langle d_{yz}|L_x|d_{x^2-y^2}\rangle$ and $\langle d_{xy}|L_x|d_{xz}\rangle$ decreases rapidly. In order to deeply understand the regulation of biaxial strain on MAE, the projected density of states is shown in Fig. 6(d). The d-orbital electron of Ta atoms of the highest occupied states and the lowest unoccupied states are both spin-up. The MAE is contributed by the coupling of spin-up occupied states with spin-up unoccupied states (uu) and spin-down unoccupied states (ud). As the strain from -2% to 3%, the energy of the spin-up and spin-down unoccupied states both increases, so both the uu and ud of $\langle d_{xy}|L_x|d_{xz}\rangle$ decreases.

## 4. Conclusions

Using first-principles calculations, we predict that Janus TaNF monolayer are excellent ferrovalley semiconductors with large valley splitting, piezoelectric polarization, high Curie temperature and huge magnetic anisotropy. A huge valley splitting of 370 meV in the conduction band minimum for Janus TaNF monolayer is realized, resulting from the cooperation of the strong SOC effect of Ta atom and intrinsic ferromagnetism. Janus TaNF monolayer exhibit larege out-of-plane piezoelectric polarizations (0.33 pm/V) because of the broken mirror symmetry. Janus TaNF monolayer possess large MAE (4.857 meV) and high Curie temperature (231 K) owing to large $\langle d_{yz}|L_x|d_{z^2}\rangle$ and strong exchange coupling parameters. The valley splitting increases under compression biaxial strain because the composition of in-plane orbitals increases. Tensile strain increases the band gap and compressive biaxial strain can reduce the band gap, resulting the transition from semiconductor to semi-metal. Curie temperature from intrinsic 231K to 373K since exchange coupling parameters increase under only -3% biaxial strain. MAE first increases and then decreases with the increase of the lattice constant. Our work reveals the promising applications of Janus TaNF monolayer in spintronics, valleytronics and piezoelectrics.

## Acknowledgements

This work is financially supported by the National Natural Science Foundation of China (Grant No. 52073308, No. 11804395), the Distinguished Young Scholar Foundation of Hunan Province (Grant No. 2015JJ1020), the Central South University Research Fund for Innovation-driven program (Grant No. 2015CXS1035) and the Central South University Research Fund for Sheng-hua scholars (Grant No. 502033019), China Postdoctoral Science Foundation (Grant No. 2022TQ0379), State Key Laboratory of Powder Metallurgy at Central South University, and the Fundamental Research Funds for the Central Universities of Central South University. This work is carried out in part using computing resources at the High Performance Computing Center of Central South University.

## Conflict of Interest

The authors declare that they have no known competing financial interests or personal relationships that could have appeared to influence the work reported in this paper.